\documentclass[11pt,a4paper]{article}
\usepackage{amsmath}
\usepackage{xarticle}
\usepackage{setspace}
\def\paperid{}

\usepackage{multirow}
\usepackage{tikz}
\usepackage{etoolbox}
\usepackage[ruled,vlined,norelsize,\languagename]{algorithm2e}
\usepackage{kbordermatrix}
\usepackage{xspace}
\usepackage{url}

\usepackage{rotating}

\begin{document}

\title{Distance-based phylogenetic inference from typing data: a unifying view}
\author{
C\'atia Vaz\footnote{Instituto Superior de Engenharia de Lisboa, Instituto Polit\'{e}cnico de Lisboa, and INESC-ID Lisboa.} \and
Marta Nascimento\footnote{Instituto Superior T\'{e}cnico, Universidade de Lisboa.} \and
Jo\~ao A. Carri\c{c}o\footnote{Faculdade de Medicina, Instituto de Microbiologia, Instituto de Medicina Molecular, Universidade de Lisboa.} \and
Tatiana Rocher\footnote{INESC-ID Lisboa.} \and
Alexandre P. Francisco\footnote{Instituto Superior T\'{e}cnico, Universidade de Lisboa, and INESC-ID Lisboa.}
}
\date{\dateline{\today}}
\maketitle

\abstract{
Typing methods are widely used in the surveillance of infectious diseases, outbreaks investigation and studies of the natural history of an infection.
And their use is becoming standard, in particular with the introduction of High Throughput Sequencing (HTS).
On the other hand, the data being generated is massive and many algorithms have been proposed for phylogenetic analysis of typing data, addressing both correctness and scalability issues.
Most of the distance-based algorithms for inferring  phylogenetic trees follow the closest-pair joining scheme.
This is one of the approaches used in hierarchical clustering.
And although phylogenetic inference algorithms may seem rather different, the main difference among them resides on how one defines cluster proximity and on which optimization criterion is used.
Both cluster proximity and optimization criteria rely often on a model of evolution.
In this work we review, and we provide an unified view of these algorithms.
This is an important step not only to better understand such algorithms, but also to identify possible computational bottlenecks and improvements, important to deal with large data sets.
}

\maketitle

\section{Introduction}

Sequence-based typing methods are rapidly becoming the gold standard for epidemiological surveillance, continuously generating huge volumes of data~\cite{Francisco2009}.
These methods provide reproducible and comparable results, being fundamental in epidemiological and genetic studies since they provide important knowledge for the surveillance of infectious diseases and about its evolution.
Phylogenetic analyses rely on typing data in order to infer relationships between different species or populations.
This analysis is made through inference algorithms using different clustering techniques~\cite{Huson2011}.

The choice of which typing method to use depends on the epidemiological context.
These methods provide further knowledge on surveillance of infectious diseases, outbreaks investigation and the natural history of an infection~\cite{robinson2010bacterial}.
Furthermore, with the introduction of High Throughput Sequencing (HTS)~\cite{Reuter2015} technology, there have been developed new typing methods like ribossomal-Multilocus Sequence Typing (ribossomal-MLST)~\cite{Maiden1998}, Multilocus Variable Number of Tandem Repeats Analysis (MLVA)~\cite{MLVA} or analysis of Single Nucleotide Polymorphisms (SNPs) through comparison against a reference genome.
Moreover, these recent advances and the resulting decrease in costs that allow the analysis of huge datasets, created the need of developing efficient and scalable methods that are able to do large scale phylogenetic analyses.

Phylogenetic analysis aims at uncovering the evolutionary relationships between different species or strains (OTU -- Operational Taxonomic Units), to obtain an understanding of their evolution.
Phylogenetic trees are widely used to address this task and are reconstructed by several different algorithms~\cite{saitou2013introduction}.

But the choice of which algorithm to use is not trivial, and most users are not fully aware of their differences.
The focus here is on distance-based analysis of DNA sequences or typing profiles, where the Hamming distances between pairs of sequences/profiles are computed~\cite{Hamming1950}.
These distances are then subject to a distance correction that is based on some appropriate model of evolution (also known as substitution model).
The resulting distance matrix is then used with a method to reconstruct a phylogenetic tree. 

This paper surveys such phylogenetic inference methods and frames them within a common general algorithm description.
Such common description allows us to reason about different algorithms based on the same premises.
We begin with a concise introduction to phylogenetic inference, clustering, and some of the main methods used to infer phylogenetic trees.
Different classifications and categorisations are then presented for those methods along their formalisation.
Finally, the problem of missing data is discussed.

\section{Phylogenetic inference}

These relationships are discovered through phylogenetic inference methods that evaluate observed heritable traits, 
such as DNA sequences or morphology under a model of evolution. These models try to describe the evolution process of the 
species from which a sequence of symbols changes into another set of traits and differ in terms of the parameters used to 
describe the rates at which one nucleotide replaces another during evolution. For instance, they are used during the 
calculation of likelihood of a tree (in Bayesian and Maximum Likelihood approaches to tree estimation) or to estimate the 
evolutionary distance between sequences from the observed differences. This enables us to infer evolutionary events that
happened in the past, and also provides more information about the evolutionary processes operating on sequences. 

Models of evolution can be classified as neutral, independent and finite-sites~\cite{hein2004}. A model is considered 
\textit{neutral} if the majority of evolutionary changes are caused by genetic drift (dominated by random processes) 
rather than natural selection. This means that the change in the frequency of a site in a population is due to random 
sampling of organisms instead of being related with its beneficial or deleterious effect. The most commonly used neutral 
models are infinite alleles model~\cite{kimura1964} and the stepwise mutation model~\cite{ohta1973}. 
They can also be considered \textit{independent} if changes in one site do not affect the probability of changes in another 
site, and \textit{finite-site} if there are finitely many sites, and so over evolution, a single site can be changed 
multiple times independently of each other. The simplest and most well-known finite sites model is the Jukes-Cantor
model~\cite{jukes1969} where all positions are equally likely to mutate and the mutant is chosen with equal probability among 
the three possible nucleotides. This model was later modified by Kimura~\cite{kimura1980} to accommodate the fact that 
transition events ($A \leftrightarrow T$ and $C \leftrightarrow G$) occur at a faster rate than all other events. 
Both models are unrealistic in the sense that all nucleotides are expected to occur with the same frequency in a random 
sequence, which is not likely to be the case for any sequence. For that reason, more sophisticated models have been introduced 
to account for subtle differences in substitution rates (\textit{e.g.} Felsenstein~\cite{felsenstein1981}, 
Hasagawa~\cite{hasegawa1985}, etc.).

We also can impose a process of mutation on top of the these models, and independently of the model, by assigning a mutation 
event with probability $\mu$ on each gene.  For instance, under the finite sites model a site is chosen randomly and a mutation
 occurs in that position according to a defined mutation process~\cite{hein2004}.

These models can assume a \textit{molecular clock} if the expected number of substitutions is constant regardless 
of which species' evolution is being examined and can be \textit{time-reversible} if it does not care which sequence is the 
ancestor and which is the descendant so long as all other parameters (\textit{e.g.} number of substitutions) are held constant. 

There are several methods that construct phylogenetic trees and they can all be seen as clustering methods because they apply 
several clustering techniques in their approach. Note also that the goal of phylogenetic analyses is to discover relationships 
between species or populations by grouping them based on some similarity criterion that underlies some model of evolution. 

These methods can be classified by the type of data they use to build the tree (Table~\ref{tab:classif-by-type-of-data}) or 
by the type of tree search algorithm (Table~\ref{tab:class-by-type-of-tree-search})~\cite{saitou2013introduction}.

\subsection{Classification by type of data}
Phylogenetic trees can be mainly built using either distance-based methods or character-state methods.
\textit{Distance-based} methods infer the relationship between individuals as
the number of genetic differences between pairs of sequences, whereas an array
of character states is used in \textit{character-state} methods. Among several
other alternatives, nucleotide sequences and amino acid sequences are used
often as character states for phylogeny construction~\cite{Stewart2003}.  In
Table~\ref{tab:classif-by-type-of-data} we can observe some examples of methods
according to this classification however the focus of this work is on distance-based
methods. Although it is also out of the scope of this paper, we should also 
note that phylogenetic trees can be built using other methodologies, such as through
the combination of phylogenetic trees built for individual partitions, e.g., individual genes~\cite{bininda2014introduction,zhang2018astral}.

\begin{table*}
\centering
\caption{Methods classification by type of data. Note that Minimum Deviation is also known as Fitch and Margoliash's method.}
\label{tab:classif-by-type-of-data}
\begin{tabular}{| c | c | c |}
\hline
Classification & Data   & Method  \\
 \hline
\multirow{3}{*}{Distance matrix}&\multirow{3}{4.1cm}{$n(n-1)/2$ pairwise distances for $n$ OTUs}&UPGMA, WPGMA, Neighbor-Joining,\\
                                &                                                               &Minimum Deviation, Minimum Evolution,\\
                                &                                                               &Distance Parsimony, goeBURST, ...\\
\hline
\multirow{2}{*}{Character-state} & \multirow{2}{4.1cm}{Array of character states (e.g. nucleotide sequences)} & Maximum Parsimony, Maximum Likelihood, \\
   &  & Bayesian, ... \\
 \hline
\end{tabular}
\end{table*}

Distance-based methods compute a distance matrix $D$ and then a phylogenetic tree $\mathcal{T}$~\cite{Pardi2016}. 
This matrix $D$ can be built directly from the pairwise distance between two sequences or from exposing them to 
a distance correction according to some model of evolution.

The simplest approach to compute $D$ is to use the \textit{normalized} Hamming distance $H(A, B)$, defined as the 
proportion of positions at which two aligned sequences A and B differ.
This distance does not take into consideration the number of back mutations 
and multiple mutations that occurred at the same position, and therefore it underestimates the true evolutionary distance. 
To rectify this, a correction formula based on some model of evolution is often used~\cite{Huson2011}.
For example, in the case of the Jukes-Cantor model, it is assumed that all substitutions are independent, and that 
 equal base frequencies, equal mutation rates and no insertions or deletions have occurred~\cite{jukes1969}.
Given the proportion of positions that differ between two sequences $A$ and $B$, $H(A, B)$, the Jukes-Cantor 
estimate of the evolutionary distance (in terms of the expected number of changes) between two sequences is given by
\begin{equation} \label{jc-model}
JC(A,B)=-\frac{3}{4}\ln\left(1-\frac{4}{3}H(A, B)\right).
\end{equation}
So, if we assume that a given set of sequences evolved according to the Jukes-Cantor model of evolution,
 to compute a distance matrix that approximates the true evolutionary distances, we first determine the \textit{normalized} 
 Hamming distance between any two sequences $A$ and $B$ in that set, and then we apply this transformation to get a corrected value.

An artificial example of a matrix $D$, built directly from the pairwise distance (\textit{unnormalized} Hamming distance), 
is presented in Equation~\ref{matrix:dist}, where each entry represents the pairwise distance, $D_{ij}$, between elements $i$ and 
$j$ where each element is an OTU (\textit{i.e.} belong to a species or population). It is easy to understand that this matrix 
is symmetric because $D_{ij} = D_{ji}$.

\patchcmd{\bordermatrix}{\left(}{\left[}{}{}
\patchcmd{\bordermatrix}{\right)}{\right]}{}{}
\begin{equation}
\label{matrix:dist}
\mathbf{D} = 
\bordermatrix{ 
	& A & B & C & D & E \cr
	A & 0 & 2 & 7 & 7 & 6 \cr
	B & 2 & 0 & 7 & 7 & 6 \cr
	C & 7 & 7 & 0 & 5 & 5 \cr
	D & 7 & 7 & 5 & 0 & 3 \cr
	E & 6 & 6 & 5 & 3 & 0}
\end{equation}

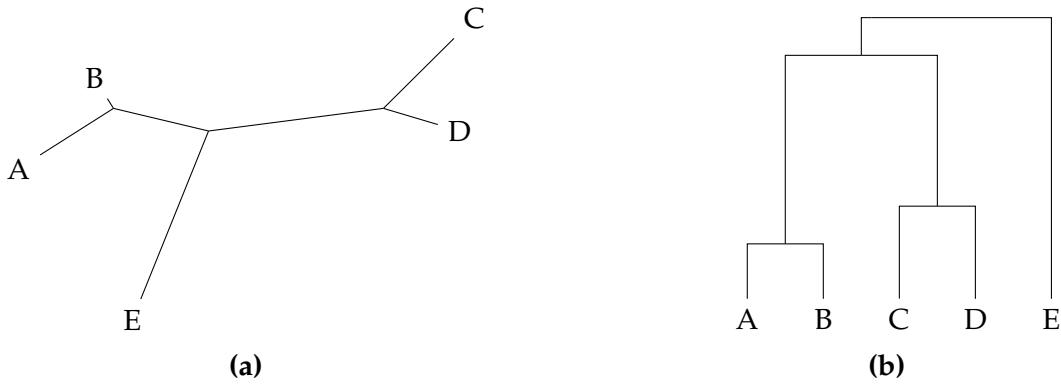
\begin{figure*}[t]
	\centering
		\begin{tikzpicture}
		\node (a)  at (0,3) {A};
		\node (b) at (1,4.2) {B};
		\node (c) at (6,5) {C};
		\node (d) at (5.8,3.5) {D};
		\node (e) at (1.5,1) {E};
		\node (ab) at (1.25,3.8) {};
		\node (cd) at (4.8,3.8) {};
		\node (abcd) at (2.5,3.5) {};
		\draw (a)-\\(ab.center);
		\draw (b)-\\(ab.center);
		\draw (c)-\\(cd.center);
		\draw (d)-\\(cd.center);
		\draw (ab.center)-\\(abcd.center);
		\draw (cd.center)-\\(abcd.center);
		\draw (e)-\\(abcd.center);
		\end{tikzpicture}
		\label{fig:phyl-trees-nj} 
		\hspace{2cm}
		\begin{tikzpicture}
		\node (a) at (0,0) {};
		\node (a) at (6,0) {};
		\node (a) at (1,0) {A};
		\node (b) at (2,0) {B};
		\node (c) at (3,0) {C};
		\node (d) at (4,0) {D};
		\node (e) at (5,0) {E};
		\node (ab) at (1.5,1) {};
		\node (cd) at (3.5,1.5) {};
		\node (abcd) at (2.5,3.5) {};
		\node (all) at (2.5,4) {};
		
		\draw  (a) |- (ab.center);
		\draw  (b) |- (ab.center);
		\draw  (c) |- (cd.center);
		\draw  (d) |- (cd.center);
		\draw  (cd.center) |- (abcd.center);
		\draw  (ab.center) |- (abcd.center);
		\draw  (abcd.center) |- (all);
		\draw  (e) |- (all);
		
		
		

		\end{tikzpicture}
		\\
		\centering \textbf{(a)}\hspace{0.5\textwidth}\textbf{(b)} \\
		\hfill \\
		\label{fig:phyl-trees-upgma} 
	\caption{Examples of phylogenetic trees: (a) an unrooted tree built by the Neighbor-Joining
method and (b) a rooted tree, or dendrogram, built by the UPGMA method.} 
	\label{fig:phyl-trees} 
\end{figure*}

The phylogenetic tree for a given distance matrix depends on the chosen
reconstruction method and on the underlying optimization criteria.
Figure~\ref{fig:phyl-trees} shows two artificial examples of different types of
trees built by two different methods in which the first uses a correction
formula based on the minimum evolution criterion (see 
Section~\ref{sub-sec-me}) and the latter assumes a molecular clock.

Phylogenetic trees can be additive or ultrametric. 
A tree $\mathcal{T}$ is considered to be \textit{additive} if every distance between two different elements 
in the tree, $D^\mathcal{T}_{ij}$, reflects the exact distance in the matrix (\textit{i.e.} $D^{\mathcal{T}}_{ij} = D_{ij}$). 
This means that the result of summing all branch lengths in the tree $\mathcal{T}$ between elements $i$ and $j$ needs to 
be the exact value present in the matrix. An \textit{ultrametric} tree, Figure~\ref{fig:phyl-trees}~(b), is an additive tree in 
which all elements are equidistant from the root. For this reason, methods that produce this kind of tree require a constant 
mutation rate over all sequences (\textit{i.e.}  the rate at which various types of mutations occur over time must 
be constant).

\subsection{Classification by type of tree search algorithm}
As mentioned before, different methods can build different trees. 
Therefore, we can classify those trees by the strategy that each method uses to find the best tree topology: 
exhaustive search methods, completely bifurcating tree search methods and stepwise clustering methods~\cite{SaitouClassification}. Some examples of these methods 
are presented in Table~\ref{tab:class-by-type-of-tree-search}.

The strategy used by \textit{exhaustive search} methods is to examine all, or a large number of possible tree topologies, 
and choose the best one according to a certain criterion.
However, when having hundreds of sequences, these methods will not be able to compute and compare all possible topologies 
within a reasonable amount of time.
Note that, for $n$ OTUs, we have as many as $(2 n - 5)!/\left(2^{(n-3)}(n-3)!\right)$ bifurcating unrooted trees, for $n>2$, or 
$(2 n -3)!/\left(2^{(n-2)}(n-2)!\right)$ bifurcating rooted trees, for $n>1$~\cite{felsenstein1978number,saitou2013introduction}.
Taking that into consideration, a new strategy called \textit{completely bifurcating tree search} method was developed 
where the number of tree topologies is limited~\cite{SaitouBifurcatingClassification}, with the number of explored topologies being indeed polynomially limited with respect to $n$.
Several measures are computed for a limited number of tree topologies to find the best tree in methods belonging to this category. Examples of completely bifurcating tree search methods are minimum evolution method and minimum deviation method.

\begin{table*}[t]
\centering
\caption{Method classification by tree search strategy.}
\label{tab:class-by-type-of-tree-search}
\begin{tabular}{| c | c | c |}
\hline
\multicolumn{2}{|c}{Classification} & Method  \\
 \hline
\multicolumn{1}{|c|}{\multirow{5}{3cm}{\centering Exhaustive search}} & \multicolumn{1}{c|}{\multirow{2}{4cm}{\centering Standard}} & Minimum Deviation \\&& Minimum Evolution\\ 
\cline{2-2} 
	&  \multicolumn{1}{c|}{\multirow{3}{4cm}{\centering Completely bifurcating tree search}} & Maximum Parsimony\\&& Maximum Likelihood \\&& ... \\
\hline
\multicolumn{2}{|c|}{\multirow{7}{*}{Stepwise clustering}} & Neighbor-Joining \\
	\multicolumn{2}{|c|}{} & UPGMA \\
	\multicolumn{2}{|c|}{} & WPGMA \\
	\multicolumn{2}{|c|}{} & Single-Linkage \\
	\multicolumn{2}{|c|}{} & Complete-Linkage \\
	\multicolumn{2}{|c|}{} & ``Wagner distance'' \\
	\multicolumn{2}{|c|}{} & ...\\
 \hline
\end{tabular}
\end{table*}

The minimum deviation method uses the number of differences between observed and estimated distances 
(``percent standard deviation'') as criterion of choosing the best tree topology~\cite{FitchandMargoliash}. 
The estimated distance between a pair of OTUs for a given tree topology is obtained by summing all branches connecting 
these two OTUs.

Another criterion is the minimisation of the sum of branch lengths (principle of minimum evolution).
This principle is used in the minimum evolution method where the tree topology which has the smallest sum of 
branch lengths is searched~\cite{EdwardsCavalli1965}.
Neighbor-joining (NJ) method also uses this principle of minimum evolution (by using least-squares approach for 
estimating the sum of branch lengths), although it is a stepwise clustering method while the minimum evolution 
method is a completely bifurcating tree search method.
The basic procedure of this method is first to obtain the neighbor-joining tree (Saitou and Nei~\cite{Saitou1987})
 and then to search for a tree with the minimum value of the sum of branch lengths by examining all trees that are 
 closely related to the neighbor-joining tree~\cite{Rzhetsky1992}.
 
 The maximum parsimony method is related to the principle used in minimum evolution methods 
since it tries to minimize (or maximize parsimony) the number of changes on character states, 
such as nucleotide or amino acid sequences, on the given tree topology~\cite{Camin1965}. 
The maximum likelihood method uses standard statistical techniques for inferring probability distributions 
to assign probabilities to particular possible phylogenetic trees~\cite{felsenstein1981}.

\textit{Stepwise clustering} methods recursively examine a local relationship between different OTUs and find the best one.
Most of distance-based methods use this strategy as we can see in Table \ref{tab:class-by-type-of-tree-search}.

\section{Distance-based methods}
Distance-based methods take as input the pairwise distances for a set of OTUs, represented by a matrix $D$, 
possibly corrected according to some model of evolution. The actual tree is then computed from the distance matrix by 
running a clustering algorithm that starts with the most similar sequences (Globally Closest Pair) or by trying to 
minimize the total branch length of the tree (Minimum Evolution Principle). Table~\ref{tab:clustering-algorithms} provides a 
categorisation of most well known phylogenetic inference methods. Most of them are agglomerative methods, and two of them are taxon additive, namely the GME and BME methods~\cite{Desper2002}.
These are addition algorithms that at each step insert a new element to a growing tree, but they select the best position in the tree to add the new element instead of performing agglomerative steps.
All  other methods in Table~\ref{tab:clustering-algorithms-2} are agglomerative methods and are variations of 
Algorithm~\ref{alg:hc-scheme}.

An important property to be considered while discussing distance-based methods is
that of {\em statistical consistency}. Let $\mathcal{T}$ be the correct tree,
$D^{\mathcal{T}}$ the associated distance matrix, and $D$ the matrix of estimated distances.
If $D$ is a consistent estimate of $D^{\mathcal{T}}$, then the more data we have, the closer $D$ is to 
$D^{\mathcal{T}}$.
Hence, a method is said to be statistically consistent if the probability of getting the correct
tree $\mathcal{T}$ increases as $D$ converges to $D^{\mathcal{T}}$~\cite{desper2004theoretical,Pardi2016}.

%
\begin{table*}
\centering
\caption{Phylogenetic inference methods and properties categorized by optimization criterion. The time is expressed in function of the number of profiles, and thus we are not taking into account the length of the profile. Time $t$ in NJMerge pipeline is the time to estimate species subtrees.}
\label{tab:clustering-algorithms}
\begin{tabular}{|c|c|c|c|c|}
\hline
 Criterion & Method & \multirow{1}{2cm}{\centering Tree} & {Complexity} & Proposed by \\ 
\hline
		\multicolumn{1}{|c|}{\multirow{6}{2cm}{\centering Globally Closest Pairs (Molecular-Clock)}} & UPGMA & \multirow{5}{2cm}{\centering Ultrametric; Rooted (\textit{e.g.} dendrogram)}& {$O(n^2)$}  & Sokal and Michener~\cite{Sokal1958}\\
		\multicolumn{1}{|c|}{} & WPGMA &  & {$O(n^2)$} & Sokal and Michener~\cite{Sokal1958}\\ 
		\multicolumn{1}{|c|}{} & SL &  & {$O(n^2)$} & Sibson~\cite{slink}\\ 
		\multicolumn{1}{|c|}{} & CL &  & {$O(n^2)$} & Dafays~\cite{clink}\\
		\multicolumn{1}{|c|}{} & UPGMC &  & {$O(n^2)$} & Sneath and Sokal~\cite{sneath1973}\\
		\multicolumn{1}{|c|}{} & WPGMC &  & {$O(n^2)$}  & Sneath and Sokal~\cite{sneath1973}\\ \hline
		\multicolumn{1}{|c|}{\multirow{15}{2cm}{\centering Minimum Evolution}} & NJ & \multirow{17}{2cm}{\centering Additive; Unrooted} & {$O(n^5)$} & Saitou and Nei~\cite{Saitou1987} \\
		\multicolumn{1}{|c|}{} & NJ &   & {$O(n^3)$} & Studie and Kepler~\cite{Studier}\\ 
		\multicolumn{1}{|c|}{} & BIONJ  &  & {$O(n^3)$}   &Gascuel~\cite{GascuelBIONJ1997}\\ 
		\multicolumn{1}{|c|}{} & UNJ &   & {$O(n^3)$}  &Gascuel~\cite{gascuel1997concerning}\\ 
		\multicolumn{1}{|c|}{} & RapidNJ &   &{$O(n^3)$}  &Simonsen, {\em et al.}~\cite{Simonsen2008}\\ 
		\multicolumn{1}{|c|}{} & ERapidNJ &   & {$O(n^3)$}  &Simonsen, {\em et al.}~\cite{Simonsen2010}\\ 
		\multicolumn{1}{|c|}{} & QuickTree &   & {$O(n^3)$}  &Howe, {\em et al.}~\cite{Howe2002}\\ 
		\multicolumn{1}{|c|}{} & QuickJoin &   &  {$O(n^3)$}&Mailund and Pedersen~\cite{Mailund2004}\\ 
		\multicolumn{1}{|c|}{} & NINJA &   & {$O(n^3)$}  &Wheeler~\cite{Wheeler2009}\\ 
		\multicolumn{1}{|c|}{} & FastJoin &   & {$O(n^3)$}  &Wang, {\em et al.}~\cite{Wang2012}\\ 
		\multicolumn{1}{|c|}{} & ClearCut (RNJ) &   &  {$O(n^3)$}&Shenman, {\em et al.}~\cite{Sheneman2006}\\ 
		\multicolumn{1}{|c|}{} & FastME &   & {$O(n^3)$} &Lefort, {\em et al.}~\cite{Lefort2015}\\
		 \multicolumn{1}{|c|}{} & FastTree &   & {$O(n \sqrt{n} \ log(n))$} &Price, {\em et al.}~\cite{price2009fasttree}\\
		  \multicolumn{1}{|c|}{} & FastTree 2 &   & {$O(n \sqrt{n} + n \ log(n))$} &Price, {\em et al.}~\cite{price2010fasttree}\\
		\multicolumn{1}{|c|}{} & FNJ &   & {$O(n^2)$}  &Elias and Lagergren~\cite{Elias2009}\\ 
		\multicolumn{1}{|c|}{} & GME &   & {$O(n^2)$}  &Desper and Gascuel~\cite{Desper2002}\\ 
		\multicolumn{1}{|c|}{} & BME &   &  {$O(n^3)$}& Desper and Gascuel~\cite{Desper2002}\\
		\multicolumn{1}{|c|}{} & goeBURST &   & {$O(n^2)$}  &Francisco, {\em et al.}~\cite{Francisco2009}\\
\multicolumn{1}{|c|}{} & FHP &  & {$O(n^2)$}  & Foulds {\em et al.}~\cite{foulds1979} \\
\multicolumn{1}{|c|}{} & NJMerge pipeline & & {$O(t+n)$} & Molloy {\em et al.}~\cite{molloy2019statistically} \\
\hline
\end{tabular}
\end{table*}

\begin{algorithm}[!t]
	\textbf{Input:} A matrix $D$ over a set of elements $S$ (OTUs). \\
	\textbf{Output:} A cluster-hierarchy $\mathcal{H}$ over $S$.\\
	\textbf{Initialisation:} Initialize the cluster-set $C$ by defining a singleton cluster $C_{i} = \{i\}$ for every element $i \in S$. Initialize output hierarchy $\mathcal{H} \leftarrow C$. \\ 
	\textbf{Loop:} While $|C| > 1$ do:
	\begin{enumerate}
		\item \textbf{Cluster-pair selection:} Select a pair of distinct clusters \{$C_i$, $C_j$\} $\subseteq C$ from $D$, according to some criterion. 
		\item \textbf{Cluster-pair joining:} Remove $C_i,C_j$ from the cluster set $C$ and replace them with $C_u = \{C_i \cup C_j\}$. Add $C_u$ to the hierarchy $\mathcal{H}$ and calculate the branch length for each element ($D_{iu}$ and $D_{ju}$).
		\item \textbf{Reduction:} Update matrix $D$ by calculating the new values $(C_{uk})$ for every $C_k \in C' \setminus \{C_i \cup C_j\}$.
	\end{enumerate}
	\textbf{Finalize:} Return the hierarchy $\mathcal{H}$.
	\caption{General scheme for hierarchical agglomerative clustering methods based on distance matrices.}
	\label{alg:hc-scheme}
\end{algorithm}

\begin{sidewaystable}
\centering
\caption{Reduction formulas of phylogenetic inference methods. 
($C_u = \{C_i \cup C_j\}$). $n$ is the number of taxa.
NJ methods  take as input a dissimilarity matrix $D$ and then create a new matrix $Q$ based on the application of 
some criterion over $D$.
FastJoin uses NJ's equation to compute distances between taxa and clustered node, and a second equation for distances between clustered nodes, with $C_v = \{C_s \cup C_t\}$. ${}^{*}$In goeBURST, in the cluster-pair selection if there is still a tie at distance 3, it is used the occurrence frequency of OTUs and lastly the assigned OTU number. ${}^{**}$BIONJ uses the variance matrix $V$ of the dissimilarity matrix for cluster pair selection.  ${}^{***}$FastME, FastTree and FastTree 2 apply topological rearrangements after executing the NJ algorithm. }
\label{tab:clustering-algorithms-2}
{\small
\begin{tabular}{|c|c|c|c|}
\hline
 Criterion & Method &  Reduction Formula & Cluster Pair Selection Criterium\\ 
\hline
		\multicolumn{1}{|c|}{\multirow{6}{2cm}{\centering Globally Closest Pairs (Molecular-Clock)}} & UPGMA & $D_{u k} = \frac{|C_i|}{|C_i| + |C_j|}D_{i k} + \frac{|C_j|}{|C_i| + |C_j|}D_{j k}$ & {\multirow{6}{3cm}{\centering smallest  pairwise distance}}  \\
		\multicolumn{1}{|c|}{} & WPGMA & $D_{u k} = \frac{1}{2}(D_{i k} + D_{j k})$  &  \\ 
		\multicolumn{1}{|c|}{} & SL & $D_{u k} = \min\{D_{i k}, D_{j k}\}$ &  \\ 
		\multicolumn{1}{|c|}{} & CL & $D_{u k} = \max\{D_{i k}, D_{j k}\}$ & \\			
		\multicolumn{1}{|c|}{} & UPGMC &  $D_{u k} = \frac{|C_i|}{|C_i| + |C_j|}D_{i k} + \frac{|C_j|}{|C_i| + |C_j|}D_{j k} - \frac{|C_i||C_j| }{(|C_i| + |C_j|)^2}D_{i j}$  & \\
		\multicolumn{1}{|c|}{} & WPGMC & $D_{u k} = \frac{1}{2}D_{i k} + \frac{1}{2}D_{j k} - \frac{1}{4}D_{i j}$ & \\ \hline
		\multicolumn{1}{|c|}{\multirow{15}{2cm}{\centering Minimum Evolution}} & NJ & $Q_{ij} = \frac{1}{2}D_{ij}+\frac{1}{2(r-2)}\sum\limits_{{}_{k \neq i,j}^{k=1,}}^{r}(D_{ik}+D_{jk})+\frac{1}{r-2}\sum\limits_{{}_{k \neq i,j}^{k=1,}}^{r}\sum\limits_{{}_{l \neq i,j}^{l=k,}}^{r}D_{kl}$ & smallest pairwise over $Q$  \\
		\multicolumn{1}{|c|}{} & NJ & $Q_{ij} = (r-2) D_{ij}-\sum\limits_{k=1, k \neq i,j}^{r}D_{ik}-\sum\limits_{k=1, k \neq i,j}^{r}D_{jk}$  & smallest pairwise value over $Q$ \\ 
		\multicolumn{1}{|c|}{} & BIONJ  & $Q_{ij} = \frac{D_{ij}}{l_s}$, where $l_s$ is the sequence length & {smallest pairwise distance over $V^{**}$}  \\ 
		\multicolumn{1}{|c|}{} & UNJ &  equal to NJ & smallest pairwise value over $Q$ \\ 
		\multicolumn{1}{|c|}{} & RapidNJ &   equal to NJ  &smallest pairwise value over $Q$   \\ 
		\multicolumn{1}{|c|}{} & ERapidNJ & equal to NJ  & smallest pairwise value over $Q$  \\ 
		\multicolumn{1}{|c|}{} & QuickTree & equal to NJ  & smallest pairwise value over $Q$  \\ 
		\multicolumn{1}{|c|}{} & QuickJoin & equal to NJ &  smallest pairwise value over $Q$\\ 
		\multicolumn{1}{|c|}{} & NINJA &  equal to NJ & smallest pairwise value over $Q$   \\ 
		\multicolumn{1}{|c|}{} & FastJoin &  NJ's equation or $Q_{uv} = (Q_{is} + Q_{js} + Q_{it} + Q_{jt}- Q_{ij} - Q_{st})$ & smallest pairwise value over $Q$ \\ 
		
 \multicolumn{1}{|c|}{} & {\multirow{3}{2cm}{ ClearCut (RNJ) }} & {\multirow{3}{4cm}{\centering equal to NJ  }} & any two nodes  after it is determined\\
 	                                    &    &   & that they are closer to \\		
		               &    &   & each other that any other node\\	
		\multicolumn{1}{|c|}{} & FastME &   equal to NJ$^{***}$ & smallest pairwise value over $Q$ \\ 	
	 	    \multicolumn{1}{|c|}{} & {\multirow{2}{2cm}{ \centering FastTree}} & {\multirow{2}{4cm}{\centering equal to NJ$^{***}$ } } & smallest pairwise value, \\
 	                                    &       &   & in a list of putative closest neighbours\\ 	
	 \multicolumn{1}{|c|}{} & {\multirow{2}{2cm}{ \centering FastTree 2} } & {\multirow{2}{4cm}{\centering equal to NJ$^{***}$ } } & smallest pairwise value, \\
 	                                    &       &   & in a list of putative closest neighbours\\ 					 
     \multicolumn{1}{|c|}{} & {\multirow{2}{2cm}{ \centering FNJ}} & {\multirow{2}{4cm}{\centering equal to NJ} } & smallest pairwise value, \\
 	                                    &       &   & occurring in visible set $V$\\ 	                                    
 \multicolumn{1}{|c|}{} & {\multirow{2}{2cm}{ goeBURST}} & {\multirow{2}{4cm}{ no update with respect to distances}} &  smallest pairwise distance over $D$ \\
 	                                    &       &   & until distance $3^{*}$\\                               
\multicolumn{1}{|c|}{} & FHP &  no update with respect to distances & {smallest pairwise distance over $D$}  \\
\multicolumn{1}{|c|}{} & NJMerge pipeline & equal to NJ & {smallest pairwise distance over $D$ and respect of subtree set topology} \\
\hline
\end{tabular}
}
\end{sidewaystable}

\subsection{Globally Closest Pairs methods}

GCP based algorithms are widely used in phylogeny. They receive as input a dissimilarity matrix containing all pairwise 
differences between elements and return a hierarchy of clusters.
First, a pair of clusters is chosen based on the minimum dissimilarity criterion over the matrix $D$.
If a tie occurs, the selection between those cluster-pairs is arbitrary.
Then, the selected pair ($C_i, C_j$) is removed from the set, joined together to form one single cluster 
($C_u = \{C_i \cup C_j\}$) and added to the hierarchy $\mathcal{H}$.
The distance between each element of the selected pair to the new cluster ($D_{iu}$ and $D_{ju}$) is set to $D_{ij}/2$.
Finally, in the reduction step, all dissimilarities from $C_i$ and $C_j$ to any other element need to be recalculated 
taking into account the new cluster $C_u$ in order to update the dissimilarity matrix.
The algorithm ends when there are no more clusters to join.

UPGMA (Unweighted Pair Group Method with Arithmetic-mean)~\cite{Sokal1958}, WPGMA (Weighted Pair Group Method with Arithmetic-mean)~\cite{Sokal1958}, 
SL (Single-Linkage)~\cite{slink} and CL (Complete-Linkage)~\cite{clink} are different variants of GCP, and they all differ on the reduction formula 
used in step 3:

\begin{itemize}
\item{UPGMA}
\begin{equation} \label{eq:upgma}
D_{u k} = \frac{|C_i|}{|C_i| + |C_j|}D_{i k} + \frac{|C_j|}{|C_i| + |C_j|}D_{j k},
\end{equation}
\item{WPGMA}
\begin{equation} \label{eq:wpgma}
D_{u k} = \frac{1}{2}(D_{i k} + D_{j k}),
\end{equation}
\item{Single-Linkage}
\begin{equation} \label{eq:sl}
D_{u k} = \min\{D_{i k}, D_{j k}\},
\end{equation}
\item{Complete-Linkage}
\begin{equation} \label{eq:cl}
D_{u k} = \max\{D_{i k}, D_{j k}\}.
\end{equation}
\end{itemize}

\subsection{Minimum Evolution principle} \label{sub-sec-me}

Clustering methods try to find the optimal tree using different heuristics. 
Some models use the minimum evolution principle, 
where a tree is considered to be optimal if it has the shortest total branch lengths. 

In order to get the minimum evolution tree it must be decided how the branch lengths are estimated and then how the tree 
length is calculated from these branch lengths. As it will be demonstrated next, NJ~\cite{Studier} simply defines the tree length 
as the sum of all branch lengths, regardless of whether they are positive or negative. In practice, branch lengths 
are usually estimated through a least-squares method.
Given a distance matrix $D$ and a fixed tree $\mathcal{T}$, if ordinary least-squares (OLS) is used, then branch
lengths are estimated by minimizing $\sum_{i=1}^n \sum_{j=1}^n (D_{i j} - D_{i j}^\mathcal{T})^2$.
Other least-squares methods can be used, namely weighted least-Squares (WLS) and generalized least-squares (GLS).
If all distance estimates can be assumed to be independent and to have the same variance, \textit{OLS} should be used. 
If distance estimates are independent but could have different variances, then \textit{WLS} or \textit{GLS} should be 
considered introducing weights to take into account such differences, with \textit{GLS} not imposing any restriction and being able to benefit from the covariances of the distance 
estimates~\cite{Desper2002}.

The neighbor-joining is the most common method and has been widely used in phylogeny inference. 
It is a greedy and agglomerative clustering algorithm, which produces a tree in a bottom-up approach by iteratively combining OTUs. 
All variants of this method try to improve on its accuracy in reconstructing the true phylogenetic tree and also by trying 
to decrease its computational cost. We describe here its most widely used variants. 

\subsubsection{Neighbor-Joining and variants} 
Neighbor-Joining algorithm is the most commonly used method in phylogenetics and several variants of this algorithm 
have been introduced over the years. While some try to optimize the formulas used by NJ to better estimate the true and 
optimal tree, others try to improve its efficiency both in terms of running time and memory usage. We will start by 
explaining NJ along with the variants that differ on the formulas they use, and then a brief description of the ones that 
seek to reduce the overall complexity.

NJ (by Studier and Kepler~\cite{Studier}), UNJ~\cite{gascuel1997concerning} , BIONJ~\cite{GascuelBIONJ1997}, FNJ~\cite{Elias2009} and ClearCut~\cite{Sheneman2006} methods are variants of NJ 
(by Saitou and Nei~\cite{Saitou1987}) that differ from the latter by applying different criteria over the three steps of the 
generalized Algorithm~\ref{alg:hc-scheme}.

These algorithms take as input a dissimilarity matrix $D$ and then create a new matrix $Q$ based on the application of 
some criterion over $D$; $Q$ is then used in the step 1 of Algorithm~\ref{alg:hc-scheme} as criterion,
being updated along $D$. This criterion relies on the minimum evolution principle.
Original NJ (by Saitou and Nei) defines $Q$ as minimizing the least-squares length estimate of the tree~\cite{gascuel2006neighbor},
with
\begin{equation}\label{eq:step1njsn}
Q_{ij} = \frac{1}{2}D_{ij}+\frac{1}{2(r-2)}\sum\limits_{{}_{k \neq i,j}^{k=1,}}^{r}(D_{ik}+D_{jk})+\frac{1}{r-2}\sum\limits_{{}_{k \neq i,j}^{k=1,}}^{r}\sum\limits_{{}_{l \neq i,j}^{l=k,}}^{r}D_{kl},
\end{equation}
where $r$ represents the number of OTUs in $D$. 
Studier and Kepler~\cite{Studier} propose replacing the Saitou and Nei criterion by
\begin{equation}\label{eq:step1njsk}
Q_{ij} = (r-2) D_{ij}-\sum\limits_{k=1, k \neq i,j}^{r}D_{ik}-\sum\limits_{k=1, k \neq i,j}^{r}D_{jk}.
\end{equation}
This new criterion has the advantage of leading to a complexity of $O(n^3)$ while the first leads to $O(n^5)$. 
These two criteria were proven to be equivalent by Vach and Degens~\cite{VachDegens1991}, with several 
implementations available~\cite{Nascimento2016}. 

FNJ method uses a similar criterion as the latter NJ (Equation \ref{eq:step1njsk}), but instead of choosing the minimum in 
$Q$ chooses from a different set (called \textit{visible set}) of size $O(n)$.  Given each taxa $i$, the pair $V_i=(i,j)$ is visible with respect to 
$Q$ if $j=argmin \{ Q_{jx}: x \in S \}$, with $S$ the set of taxa. Thus, the visible set $V$ is defined as $V = \{ V_i : i \in S \}$.
At each step $s$ of the algorithm, $V$ should be updated, {\em i.e.}
$V_{s+1} = ( V_{s} \setminus \{ (x,y) : x=i$ or $x= j \}) \cup V_u$.
%
BIONJ method uses a simple first-order model of the variances and covariances of evolutionary distance, 
$Q_{ij}=D_{ij}/l_s$, where $l_s$ represents the sequence length.
At each step it allows the selection of the pair which minimizes the variance of the dissimilarity matrix.
%
ClearCut relaxes the requirement of exhaustively searching the input matrix at each step to find the closest pair of nodes to join.
Although it determines and joins the pair of nodes that are closer to each other, it does so while matrix updates due to joins are delayed. Hence candidate nodes may not be the closest pair of nodes in NJ sense.
All the remaining variants use the same criterion defined by Studier and Kepler~\cite{Studier}, minimizing the least-squares estimate of the tree length.

Branch lengths are computed in step 2. UNJ (Equation~\ref{eq:step2unj}) differs from the remaining (Equation~\ref{eq:step2nj}) because it uses an unweighted version of the equation used by NJ, relying on OLS~\cite{gascuel2006neighbor},
\begin{equation} \label{eq:step2nj}
D_{iu}=\frac{1}{2}D_{ij}+\frac{1}{2(r-2)}\left[\sum\limits_{{}_{k \neq i,j}^{k=1,}}^{r}(D_{ik}-D_{jk})\right],\quad D_{ju}=D_{ij}-D_{iu},
\end{equation}
\begin{multline}\label{eq:step2unj}
D_{iu}=\frac{1}{2}D_{ij}+\frac{1}{2(|S|-|C_u|)}\left[\sum\limits_{{}_{k \neq i,j}^{k=1,}}^{r}|C_k|(D_{ik}-D_{jk}) \right],\quad  \\ D_{ju}=D_{ij}-D_{iu}.
\end{multline}
It is easily observed that this equation is obtained from NJ by setting $|C_k|=1$ for all $k \neq i,j$ and by replacing $(|S|-|C_u|)$ by $ \sum\limits_{k \neq i,j}^{}|C_k|$ which is then equal to $(r-2)$.

Finally, in step 3, the dissimilarity matrix is reduced by deleting a pair of OTUs ($i,j$) and by estimating new distances between the new OTU $u$ to any other OTU $k$ ($D_{uk}$) using the general reduction formula
\begin{equation}\label{eq:step3general}
D_{uk}=\lambda(D_{ik} - D_{iu}) + (1-\lambda)(D_{jk} - D_{ju}),
\end{equation}
where $D_{iu}$ and $D_{ju}$ are given by either (\ref{eq:step2nj}) or (\ref{eq:step2unj}), and $\lambda$ is the weight that each algorithm assigns to each branch.
NJ method defines $\lambda$ by giving an equal weight to both original branch lengths ($D_{ik}$ and $D_{jk}$), {\em i.e.}, $\lambda = \frac{1}{2}$.
But Saitou and Nei~\cite{Saitou1987} do not consider the newly computed branches, i.e., $D_{iu} = D_{ju} = 0$, whereas Studier and Kepler~\cite{Studier} do. UNJ defines the weight $\lambda$ as proportional to the number of OTUs contained in the clusters and hence gives the same weight to each of these OTUs, $\lambda = \frac{|C_i|}{|C_i|+|C_j|}$.
And BIONJ assigns the weight according to their variance,
\begin{equation}\label{eq:step3bionj-d}
\lambda = \frac{1}{2}+\frac{\sum\limits_{k=1, k \neq i,j}^{r}(V_{jk}-V_{ik})}{2(r-2)V_{ij}},
\end{equation}
with $\lambda \in [0,1]$.

Regarding the reduction of $Q$, all methods that use NJ criterion need to update the entire matrix using the same equations that were used at the beginning Equations \ref{eq:step1njsn} and \ref{eq:step1njsk}, while BIONJ just needs to calculate the new values between the new OTU $u$ to any other OTU $k$ $(D_{uk})$ using Equation \ref{eq:step3bionj-v}, that is identical to Equation \ref{eq:step3general}, but instead of using the distance matrix it uses the variance matrix,
\begin{equation}\label{eq:step3bionj-v}
V_{uk}=\lambda(V_{ik} - V_{iu}) + (1-\lambda)(V_{jk} - V_{ju}).
\end{equation}

FNJ method also uses the same equation as NJ, Equation \ref{eq:step3general}, but has an additional step for updating the \textit{visible set} by removing the previously selected \textit{visible pair} and adding a new \textit{visible pair} for $C_u$. 

The complexity $O(n^3)$ of neighbor-joining method is directly related with the selection and reduction steps because it joins the two clusters that are closest to each other in $O(n^2)$ and then updates the distance matrix with the new distance values, at each step, for $O(n)$ OTUs. 

Regarding neighbor-joining efficiency, several methods were developed to address running time and memory issues~\cite{Pardi2016}. These include  QuickJoin by Mailund and Pedersen~\cite{Mailund2004}. NINJA by Wheeler~\cite{Wheeler2009}. RapidNJ by Simonsen {\em et al.}~\cite{Simonsen2008}, using heuristics to find the next two elements to join avoiding the exploration of all possible pairs. ERapidNJ by Simonsen {\em et al.}~\cite{Simonsen2010}, through reducing the memory requirements of RapidNJ. FastJoin by Wang {\em et al.}~\cite{Wang2012}, decreasing the search time and the updating time of the distance matrix. QuickTree by Howe {\em et al.}~\cite{Howe2002}, making use of low data-structure overhead and using an heuristic for handling redundant data. Moreover, QuickJoin, RapidNJ and NINJA methods produce the same phylogenetic trees as NJ but improve the running time by using some techniques to find the globally minimum value of sum matrix rather than by traversing the whole sum matrix in each iteration. Although all these methods take $O(n^3)$ running time in the worst case, in practice they perform faster.

Molloy {\em et al.}~\cite{molloy2019statistically} proposed a polynomial-time extension of NJ, namely NJMerge. It uses a divide and conquer method to construct a tree from a dissimilarity matrix and a set of subtrees, where the resulting tree matches the topology of subtrees.  NJMerge can be included in a pipeline to construct the tree from the dissimilarity matrix only, where the user needs to choose a method and setup a set of parameters to build the set of subtrees. Although NJMerge can fail to compute the tree (in $1\%$ of the test cases), the method can improve the accuracy of NJ method, and speedup the execution time for computing trees from multi-locus datasets of up to a thousand species.

Application of different techniques for topological rearrangements in the tree are other possible optimizations.
Usually, they are used by heuristic algorithms like FastME by Lefort {\em et al.}~\cite{Lefort2015}, FastTree and FastTree2 by Price {\em et al.}~\cite{price2009fasttree, price2010fasttree}, BNNI and FASTNNI~\cite{desper2002fast}, which search for an optimal tree structure. These techniques are Nearest Neighbor Interchanges (NNI)~\cite{jiang2000computing}, Subtree Pruning and Regrafting (SPR)~\cite{whidden2018efficiently}, and Tree Bisection and Reconnection (TBR)~\cite{allen2001subtree}. All of these techniques are implemented seeking to optimize some criterion such as
balanced minimum evolution (BME)~\cite{Lefort2015}, the OLS version of criterion (OLSME)~\cite{Lefort2015} or WLS version of criterion (WLSME)~\cite{desper2004theoretical}. It is important to notice that Desper and Gascuel~\cite{desper2004theoretical} demonstrated that the  balanced minimum evolution  is a special case of the WLS approach, with biologically meaningful variances of the distance estimates.
Price {\em et al.} obtained a subquadratic execution time in FastTree2 by receiving as input the OTU representations, such as sequences, instead of the distance matrix, and using heuristics to estimate distances and limit the number of NNI and SPR moves: for example, FastTree2 searches for SPR moves in each subtree twice instead of until convergence.
Nevertheless, and regardless of the implemented optimization technique, their ultimate goal is always to efficiently obtain the most accurate tree.
In what concerns
statistical consistency, it has been shown that minimum evolution principle combined with OLS as well as the balanced minimum evolution principle are consistent~\cite{rzhetsky1993theoretical,desper2004theoretical}.

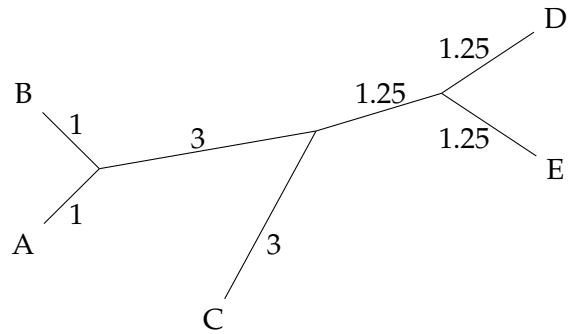
\begin{figure}
	\centering
	\begin{tikzpicture}
	\node (a)  at (-1,2) {A};
	\node (b) at (-1,4) {B};
	\node (ab) at (0,3) {};
	\node (d) at (6,5) {D};
	\node (e) at (6,3) {E};
	\node (de) at (4.5,4) {};
	\node (c) at (1.5,1) {C};	
	\node (abcd) at (2.85,3.5) {};
	
	\node (1) at (-0.30,3.6) {1};	
	\node (1) at (-0.30,2.4) {1};	
	\node (2) at (4.8,4.6) {1.25};	
	\node (2) at (4.8,3.4) {1.25};	
	\node (2) at (3.7,4) {1.25};	
	\node (3) at (1.3,3.4) {3};	
	\node (4) at (2.3,2) {3};	
	
	\draw (a)-\\(ab.center);
	\draw (b)-\\(ab.center);
	\draw (d)-\\(de.center);
	\draw (e)-\\(de.center);
	\draw (ab.center)-\\(abcd.center);
	\draw (de.center)-\\(abcd.center);
	\draw (c)-\\(abcd.center);
	\end{tikzpicture}
	\caption{Phylogenetic tree obtained with NJ (Studier and Kepler~\cite{Studier}) for matrix~\ref{matrix:dist}.} 
	\label{fig:nj-example} 
\end{figure}

Figure~\ref{fig:nj-example} presents an example of a  phylogenetic tree obtained with NJ with the criterion by Studier and Kepler~\cite{Studier}.

\subsubsection{GME and BME}
\textit{GME} and \textit{BME} addition algorithms
are stepwise greedy algorithms that at each step insert a new element to a
growing tree~\cite{Desper2002}. This approach differs from the last ones
because instead of having a selection criterion for choosing the best element to
be added in a fixed position in the tree, it has a selection criterion for
choosing the best position in the tree to add any element. These algorithms add
iteratively every element into the tree by minimizing the cost of insertion
that is calculated for every edge in the tree. The scheme for both of these is
presented in Algorithm~\ref{alg:gme_bme} and they both try to obtain the minimum
evolution tree by minimizing the total tree length.

While GME algorithm relies on OLS enabling it to be faster than the most common
minimum evolution based algorithms, BME is similar and related to WLS
fitting~\cite{desper2004theoretical}, being more appropriate but slower than
GME. In BME, sibling subtrees have equal weight, as opposed to the GME, where
the weight of a subtree is equal to the number of OTUs it contains. Note that
BME can be seen as a weighted version of GME, just like NJ is the weighted
version of UNJ, and WPGMA is the weighted version of UPGMA.
In what concerns
statistical consistency, both GME and BME were shown to be
consistent~\cite{rzhetsky1993theoretical,desper2004theoretical,huson2010phylogenetic}

\begin{algorithm}
	\textbf{Input:} A dissimilarity matrix $D$ between distant-2 subtrees $S$ (\textit{i.e.} if $A$ and $B$ are two disjoint subtrees, with roots $a$ and $b$ respectively, we will say that $A$ and $B$ are distant-$k$ subtrees if there are $k$ edges in the path from $a$ to $b$). \hfill \\
\hspace{10.5mm} A tree $\mathcal{T}_k$ containing three subtrees. \hfill \\
	\hspace{10.5mm} $k$, number of already processed OTUs ($k=3$). \hfill \\
	\hspace{10.5mm} An array $P$ with the number of OTUs in each subtree. \hfill \\
	\hspace{10.5mm} $n$, number of initial OTUs. \hfill \\
	\textbf{Output:} A tree $\mathcal{T}_n$ over all $n$ OTUs.\\
	\hfill \\
	\textbf{Initialization:} Initialize $k=4$.\\ 
	\hfill \\
	\textbf{Loop:} While $k < n$ do:
	\begin{enumerate}
		\item Calculate the dissimilarity between OTU $k$ and any subtree  $S$ from $\mathcal{T}_{k-1}$.
		\item Calculate the cost of inserting $k$ along edge $e$ as $f(e)$ for all edges $e$ in $\mathcal{T}_{k-1}$.
		\item Select the best edge by minimizing $f$ and insert $k$ on that edge to form $\mathcal{T}_{k}$.
		\item Update the dissimilarity matrix $D$ between every pair of distant-2 subtrees as well as the array $P$ with the number of OTUs per subtree.
	\end{enumerate}
	\textbf{Finalize:} Return the tree $\mathcal{T}_n$.
	\caption{General scheme for addition methods.}
	\label{alg:gme_bme}
\end{algorithm}

Given an ordering on the OTUs, denoted as $(1, 2, 3, \dots, n)$, for $k=4$ to $n$, a tree $\mathcal{T}_k$ is created. This is done by testing each edge of $\mathcal{T}_{k-1}$ as a possible insertion point for $k$ and comparing them using the OLS (Ordinary Least Squares) minimum evolution criterion in GME and the balanced minimum evolution criterion in BME. Then, $k$ is inserted on edge $e$ of $\mathcal{T}_{k-1}$, that edge $e$ is removed and the length of every other already existing edge is changed. After this, it is required the computation of the length of the three newly created edges. 

The \textit{tree length} $l(\mathcal{T})$ is defined as the sum of all edge lengths $l(e)$ of $\mathcal{T}$. In GME, the minimum evolution tree is an OLS minimum evolution tree that minimizes the length of the tree $l(\mathcal{T})$ and where $\mathcal{T}$ has the OLS edge length estimates. In BME, the definition for tree length holds but instead of using OLS distance it uses a balanced distance between subtrees. Given two subtrees $S_i$ and $S_j$, if they are subtrees with only one OTU ($S_i = \{i\}$ and $S_j = \{j\}$) then the dissimilarity comes directly from the distance matrix. Otherwise, if $S_i = \{i\}$ and $S_j=\{S_{j_1},S_{j_2}\}$ the dissimilarity can be computed by Equation~\ref{eq:step1gme} for GME or Equation~\ref{eq:step1bme} for BME because under a balanced weighting scheme, sibling subtrees are assigned equal weights, while in the OLS scheme the weight of a subtree is equal to the number of OTUs it contains. Therefore, the distance is also generalized to internal nodes,
\begin{equation}\label{eq:step1gme}
D_{ij} = \frac{|S_{j_1}|}{|S_j|}D_{ij_1} + \frac{|S_{j_2}|}{|S_j|}D_{ij_2},
\end{equation}
\begin{equation}\label{eq:step1bme}
D_{ij} = \frac{1}{2}(D_{ij_1} + D_{ij_2}).
\end{equation}

Suppose that we have a tree $\mathcal{T}$ with four subtrees $A$, $B$, $C$ and $D$, with $e$ is an internal edge that connects subtrees $A$ and $B$ with subtrees $C$ and $D$, and that for any subtree $K$ the distance between its root $k$ and that subtree is $D_{kK}$. In the following, the lowercase letter will  identify the root of the subtree denoted by corresponding uppercase letter. Then the tree length can be computed recursively by applying
\begin{multline}\label{eq:tree-length-all}
l(\mathcal{T})  =  \frac{1}{2}[\lambda(D_{AC}+D_{BD})+(1-\lambda)(D_{AD}+D_{BC}) +D_{AB} \\ +D_{CD}]  +  l(A) + l(B) + l(C) + l(D) - D_{aA} - D_{bB} - D_{cC} - D_{dD}.
\end{multline}

Now, in order to compute the length of an edge $e$ it is necessary to distinguish when $e$ is an internal or external edge. First, let us assume that $e$ is an internal edge that connects subtrees $A$ and $B$ with $C$ and $D$, with roots $a$, $b$, $c$ and  $d$, respectively. Then, the length estimate of $e$ can be defined by Equation \ref{eq:edge-internal-length-all}. If we consider $e$ an external edge that connects subtrees $A,B$ with a single OTU $k$, we need to apply Equation \ref{eq:edge-external-length-all}.

\begin{equation}\label{eq:edge-internal-length-all}
l(e)=\frac{1}{2}[\lambda(D_{Ak}+D_{BD})+ (1-\lambda)(D_{AD}+D_{Bk})-D_{AB}-D_{kD}].
\end{equation}
\begin{equation}\label{eq:edge-external-length-all}
l(e) = \frac{1}{2}(D_{Ak}+D_{Bk}-D_{AB}).
\end{equation}

For the GME version, $\lambda$ is replaced in Equations \ref{eq:tree-length-all} and \ref{eq:edge-internal-length-all} by $\lambda$ as in Equation \ref{eq:length-gme}. This demonstrates an important property of OLS edge length estimation because the length estimate of any given edge does not depend on the topology of the subtrees $A$, $B$, $C$ and $D$, but only on the number of OTUs contained in those subtrees.

In the balanced version of minimum evolution, $\lambda$ is replaced by $1/2$ (Equation \ref{eq:length-bme}) to assign equal weights to all subtrees, regardless of the number of OTUs they contain. The edge length estimates now depend on the topology of the subtrees, simply because the balanced distances between these subtrees depend on their topologies.

\begin{equation}\label{eq:length-gme}
\lambda = \frac{|A||D|+|B||C|}{(|A|+|B|)(|C|+|D|)}.
\end{equation}
\begin{equation}\label{eq:length-bme}
\lambda = \frac{1}{2}.
\end{equation}

Starting with GME algorithm, on step 1, the dissimilarity between OTU $k$ and any other subtree can be computed by applying \ref{eq:step1gme} recursively. On step 2, suppose that a tree $\mathcal{T'}$ is formed by moving the insertion of $k$ from edge $e_1$ to the edge $e_2$, where $e_2$ is a sibling edge of $e_1$ in $\mathcal{T}$. The length of $\mathcal{T'}$ is computed by \ref{eq:tree-length-all} as $l(\mathcal{T'})=L+f(e)$, where $L$ is the length of tree $\mathcal{T}$ and $f(e)$ depends on the computations for both $\mathcal{T}$ and $\mathcal{T'}$. Then, the process of searching for every edge $e$ of $\mathcal{T}_{k-1}$ continues by recursively moving from one edge to its neighboring edges, to obtain the cost that corresponds to the length of the tree $\mathcal{T}_{k-1}$ plus $k$ inserted on $e$. Moreover, this cost can be written as $L+f(e)$ and for the first considered edge it is sufficient to minimize $f(e)$ with $f(e)=0$. Finally, on the last step, Equation \ref{eq:step4gme} is used to update the distance between any pair $A,B$ of distant-2 subtrees, if $k$ is inserted in the subtree $A$.

\begin{equation}\label{eq:step4gme}
D_{(k\cup A)B} = \frac{1}{1+|A|}D_{kB} + \frac{|A|}{1+|A|}D_{AB}.
\end{equation}

The main difference between GME and BME is that in the latter the update can no longer be achieved using a fast method as expressed by \ref{eq:step4gme}, because the balanced distance between $k\cup A$ and $B$ now depends of the position of $k$ within $A$.

\begin{figure}
	\centering
\begin{tikzpicture}
	\node (a)  at (-1,2) {A};
	\node (b) at (-1,4) {B};
	\node (ab) at (0,3) {};
   \node (d) at (6,5) {D};
	\node (e) at (6,3) {E};
	\node (de) at (4.5,4) {};
	\node (c) at (1.5,1) {C};	
	\node (abcd) at (2.85,3.5) {};
	
	\node (1) at (-0.30,3.6) {1};	
	\node (1) at (-0.30,2.4) {1};	
	\node (2) at (4.8,4.6) {1.75};	
	\node (2) at (4.8,3.4) {1.25};	
	\node (2) at (3.7,4) {0.75};	
 \node (3) at (1.3,3.4) {3.25};	
	\node (4) at (2.5,2) {2.75};	
	
	\draw (a)-\\(ab.center);
	\draw (b)-\\(ab.center);
	\draw (d)-\\(de.center);
	\draw (e)-\\(de.center);
	\draw (ab.center)-\\(abcd.center);
	\draw (de.center)-\\(abcd.center);
	\draw (c)-\\(abcd.center);
	\end{tikzpicture}
	\caption{BME phylogenetic tree according to matrix \ref{matrix:dist}.} 
	\label{fig:bme-example} 
\end{figure}
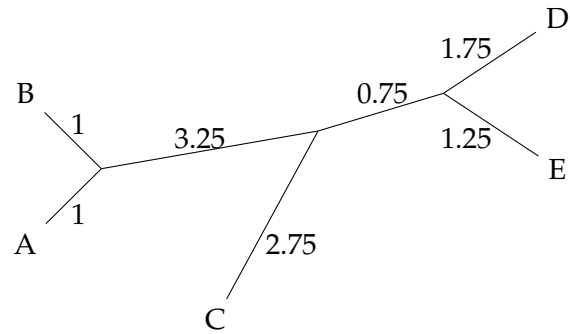

Despite of BME algorithm requiring $O(n^3)$ operations on the worst-case, in
practice it needs only $O(n^2\log n)$~\cite{Desper2002}. Therefore, both GME
and BME algorithms are faster than NJ-like algorithms which require $O(n^3)$
operations, allowing trees with thousands of elements to be constructed in few
minutes. Also, the balanced minimum evolution appears much more appropriate for
phylogenetic inference than the ordinary least-squares version (GME). This is
likely due to the fact that it gives less weight to the topologically long
distances (\textit{i.e.} those containing numerous edges), while the OLS method
puts the same confidence on each distance, regardless of its length. 
Furthermore, the balanced algorithms
produced output trees with better topological accuracy than those from NJ.
Still, NJ continues to be the most used and preferred algorithm by the
community because it guarantees that the reconstructed tree will be the correct
tree if the distance matrix is ``nearly additive''~\cite{Atteson1997}. In
practice the distance matrix rarely satisfies this condition, but neighbor
joining often constructs the most correct tree topology
anyway~\cite{Mihaescu2009}.

An example of BME phylogenetic tree according to the distance matrix in Equation \ref{matrix:dist} is presented in Figure \ref{fig:bme-example}.

\subsubsection{Distance-based MST-like methods}

As we have seen earlier, the theory of evolution predicts that existing biological species have been linked in the past by common ancestors and their relationships can be depicted as a branched diagram like phylogenetic trees. The methods that we will present here use the principle of NJ (minimum evolution) to better estimate the true phylogenies, however they were developed following a graph theoretic approach. 


The problem of determining the minimal phylogenetic tree is discussed in relation to graph theory where each OTU is represented by a point in the tree to be constructed and each link in the tree is associated with the changes between the species it connects. This problem is similar to one known in graph theory as the minimal spanning tree problem on trying to identify a subset of links in a graph that connects all points without any cycles and with the minimum possible total link length. 

Next, we will describe four methods, generic-MST, Foulds-Hendy-Penny, goeBURST and goeBURST Full MST that have been developed based on algorithms that solve the MST problem (\textit{e.g.} Bor\r {u}vka~\cite{boruvka1926}, Kruskal~\cite{kruskal1956}, Prim~\cite{prim1957}, etc.) to determine the minimal phylogenetic tree. 


The problem of finding a minimum spanning tree can be captured by the generic algorithm~\ref{alg:generic-mst}, which grows the minimum spanning tree one edge at a time. At each step, we determine a safe edge $(u,v)$ that we can add to $\mathcal{T}$. An edge is considered a \textit{safe} edge if and only if we can add it to $\mathcal{T}$ without creating a cycle in it. Therefore, all edges added to $\mathcal{T}$ guarantee that $\mathcal{T}$, in line 10, must be a minimum spanning tree. In the case of a tie, \textit{i.e.}, if a new vertex $u$ has two safe edges, $(u,v)$ and $(u,w)$, connecting to the tree $\mathcal{T}$ with equal weights, we must define a comparator, $cmp$, to decide which one to choose.

\LinesNumbered
\begin{algorithm}
	\DontPrintSemicolon
	\textbf{Input:} A graph $G=(V,E)$ where $V$ and $E$ are the set of vertices and edges, respectively. \\
	\hspace{10.2mm}$cmp$, edge weight comparator. \hfill \\
	\textbf{Output:} A minimum spanning tree $\mathcal{T}$. \\
	\hfill \\
	\Begin{
		$\mathcal{T} \leftarrow \emptyset$\;
		\While{$\mathcal{T}$ does not form a spanning tree}{	
			find a least-weight edge $(u,v)\in E$  according to comparator $cmp$ that is safe for $\mathcal{T}$,  \;
			$\mathcal{T} \leftarrow \mathcal{T} \cup \{(u,v)\}$\;
		}
		return $\mathcal{T}$\;
	}
	\textbf{Finalize:} Return $\mathcal{T}$.
	\caption{Generic-MST pseudocode.}
	\label{alg:generic-mst}
\end{algorithm}
\LinesNumberedHidden

There exist several greedy algorithms that elaborate on this generic method and they use a specific criterion to determine a safe edge, in line 8. Kruskal chooses always a least-weight edge in the graph that connects two distinct components. Prim chooses always a least-weight edge connecting the tree to a vertex not in the tree. And Bor\r {u}vka chooses the minimum-weight edge incident to each vertex of the graph.

Distance-based MST algorithm also elaborate on this generic method but the criterion is based on the computation of distances between vertices. This algorithm differs from the remaining on the comparator used to break the ties on the edges, while the others choose them arbitrarily.

The reason underlying why greedy algorithms are effective at finding minimum
spanning trees is that the set of forests of a graph forms a graphic matroid~\cite{Whitney1935,tutte1965}.


The proposed method of Foulds, Hendy and Penny~\cite{foulds1979} is an heuristic method of approaching the phylogenetic problem by transforming it into a Steiner problem~\cite{dreyfus1971} combined with Kruskal's method. It requires not only a distance matrix to represent the number of changes between the species' sequences but also the position in that sequence at which they differ, and which characters were changed from one sequence to the other.

At each iteration, the method identifies the shortest link joining two unconnected points, which when added individually to the existing graph do not create a cycle. Then, it tries to reduce the graph by ``coalescing" (\textit{i.e.} introducing Steiner points that do not represent any of the original species but represent instead a sequence of a possible intermediary species), where necessary. This means that after a link is selected and added to the tree it compares all the pairs of links incident there, and identify those changes that are common to the two links, choosing to coalesce on a pair with the maximal number of common changes (\textit{i.e.} that result in a largest reduction in the total length of the tree). If a cycle is present an attempt is made to break it by removing links which give the largest reduction in the total length of the tree. If there is more than one largest link, the cycle is left unbroken because it can subsequently be broken by later additions. The method terminates when all original species are connected. Although, if at the end the cycles are still unresolved, all the alternative trees are presented and the choice is left to the biologist who may wish to use some additional information.

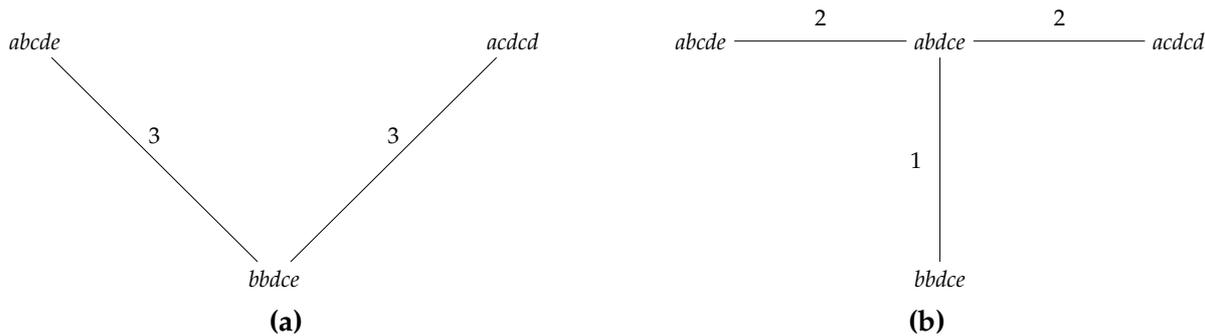
\begin{figure}
	\centering
	\resizebox{0.45\textwidth}{!}{\begin{tikzpicture}[scale=4]
		\node (1)  at (-1,1) {$abcde$};
		\node (2) at (0,0) {$bbdce$};
		\node (3) at (1,1) {$acdcd$};
		
		\node (d) at (-0.5,0.6) {3};
		\node (d) at (0.5,0.6) {3};
				
		\draw (1)-\\(2);
		\draw (2)-\\(3);
		
		\end{tikzpicture}}
	\hfill
	\resizebox{0.45\textwidth}{!}{\begin{tikzpicture}[scale=4]
		\node (1)  at (-1,1) {$abcde$};
		\node (2) at (0,0) {$bbdce$};
		\node (3) at (1,1) {$acdcd$};
		\node (4) at (0,1) {$abdce$};

		\node (d) at (-0.5,1.1) {2};
		\node (d) at (0.5,1.1) {2};
		\node (d) at (-0.1,0.5) {1};

		\draw (1)-\\(4);
		\draw (2)-\\(4);
		\draw (3)-\\(4);
		
		\end{tikzpicture}}
	\\
	\centering \textbf{(a)}\hspace{0.5\textwidth}\textbf{(b)} \\
	\hfill 
	\caption{Step by step tree representation of Foulds, Hendy and Penny (FHP) algorithm.} 
	\label{fig:fhp-example} 
\end{figure}

For instance, assume you have the following three sequences: $S_1 = abcde$, $S_2 = bbdce$ and $S_3 = acdcd$, with distances: $d(S_1,S_2)=3$, $d(S_1,S_3)=4$ and $d(S_2,S_3)=3$. We can see that $S_1$ differ from $S_2$ in the first, third and forth elements ($a\leftrightarrow b$, $c\leftrightarrow d$ and $d\leftrightarrow c$), $S_1$ differ from $S_3$ in the second, third, forth and fifth elements ($a\leftrightarrow b$, $c\leftrightarrow d$, $d\leftrightarrow c$ and $e\leftrightarrow d$) and finally, $S_2$ differ from $S_3$ in the first, second and fifth element ($b\leftrightarrow a$, $b\leftrightarrow c$ and $e\leftrightarrow d$). We now follow Kruskal, and search for the links of minimal length, not previously chosen, which, when added individually to the existing graph do not create a circuit, $d(S_1,S_2)=3$ and $d(S_2,S_3)=3$. These now give us a partial graph, illustrated in~\ref{fig:fhp-example} (a). We now attempt to reduce the graph by comparing all the pairs of links incident there, and identify those changes that are common to the two links, choosing to coalesce on a pair with the maximal number of common changes. In our example, $S_2$ is the only point with a pair of incident links and there is only one common change, ($a\leftrightarrow b$) at the first element. To represent this we introduce the Steiner point $S_4$, to represent the possible sequence: $S_4 = abdce$ (\ref{fig:fhp-example} (b)). Now, as all points are connected to the tree the algorithm is complete, so we have a spanning tree.

The present method overcomes the problem of differences in the rate of evolution on different links since all species are retained for further comparisons, even after they are linked into the graph. This is an important aspect of the method because it ensures that a particular species is included in its optimal position. Another important feature of this method is that when new ancestral species are determined, they are compared with all the existing species. This means that the links are not in a fixed position while the tree is being determined.

Note that this method does not guarantee to build a minimum spanning tree and the accuracy of that tree is very much dependent upon the suitability and reliability of the data. A single error in the sequence entries could conceivably make a significant difference to the tree if it affected closely related sequences. 

goeBURST algorithm~\cite{Francisco2009} is a globally optimized implementation of the eBURST algorithm~\cite{Feil2004} that identifies alternative patterns of descent for several bacterial species.
It implements the simplest model for the emergence of clonal complexes~\cite{Feil02012001,Smith2000} where a given genotype/OTU increases in frequency in the population, as a consequence of a fitness advantage or of random genetic drift, becoming a founder clone in the population. 
This increase is accompanied by a gradual diversification of that OTU, by mutation and recombination, forming a cluster of phylogenetically closely related strains or OTUs.
This diversification of the ``founding'' OTU is reflected in the appearance of OTUs differing only in one state when compared with the ``founding'' one; those OTUs are named single locus variants (SLV). Further diversification of those SLVs will result in the appearance of variations of the original OTU with more than one difference in the states, namely double locus variants (DLV), triple locus variants (TLV), and so on. The result is a forest, a disjoint set of trees (acyclic graphs), where each tree corresponds to a clonal complex defined for a distance of 1.

\begin{figure*}
	\centering	
	{\resizebox{0.2\textwidth}{!}{
			\begin{tikzpicture}[scale=1.5]
			\node (a) at (0,0.25) {A};
			\node (b) at (1,1) {B};
			\node (c) at (2,0.25) {C};
			\node (d) at (0.5,-0.75) {D};
			\node (e) at (1.5,-0.75) {E};
			\end{tikzpicture}}}
	\label{fig:goeburst-slv} 
	\hfill
	{\resizebox{0.20\textwidth}{!}{
			\begin{tikzpicture}[scale=1.5]
			\node (a) at (0,0.25) {A};
			\node (b) at (1,1) {B};
			\node (c) at (2,0.25) {C};
			\node (d) at (0.5,-0.75) {D};
			\node (e) at (1.5,-0.75) {E};
			
			\node (1)  at (0.4,0.8) {2};
			\draw (a)-\\(b);
			
			\end{tikzpicture}}}
	\label{fig:goeburst-dlv} 
	\hfill
	{\resizebox{0.20\textwidth}{!}{
			\begin{tikzpicture}[scale=1.5]
			\node (a) at (0,0.25) {A};
			\node (b) at (1,1) {B};
			\node (c) at (2,0.25) {C};
			\node (d) at (0.5,-0.75) {D};
			\node (e) at (1.5,-0.75) {E};
			
			\node (1)  at (0.4,0.8) {2};
			\node (2)  at (1,-0.6) {3};
			
			\draw (d)-\\(e);
			\draw (a)-\\(b);
			
			\end{tikzpicture}}}
\\
\hfill \\
\textbf{(a)}\hspace{0.37\textwidth}\textbf{(b)}\hspace{0.37\textwidth}\textbf{(c)}\\
\hfill
	\label{fig:goeburst-tlv} 
	\caption{goeBURST phylogenetic trees according to matrix \ref{matrix:dist} with different allelic distances: (a) allelic distance of one, (b) allelic distance of two and (c) allelic distance of three.} 
	\label{fig:goeburst} 
\end{figure*}
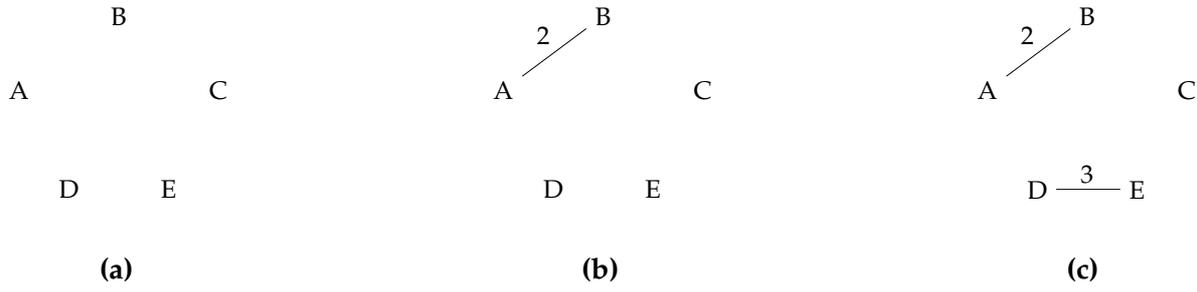

This algorithm can be stated as finding the maximum weight forest or, depending on weight definition, as finding the minimum spanning tree. Therefore, the optimal solution can be provided by a greedy approach on identifying the optimal forest with respect to the defined partial order on the set of links between STs. Due to its desirable properties and ease of implementation, goeBURST uses Kruskal algorithm to achieve that goal.

This algorithm consists of building a spanning forest in a graph where each OTU is a node and two OTUs are connected if and only if they are at distance one (\textit{i.e.} if they only are SLVs). Since this forest should be optimal with respect to link selection, the links between OTUs with higher number of SLVs must be selected. In case of a tie it should be considered the number of OTUs that are at distance two (\textit{i.e.} number of DLVs), the number of OTUs that are at distance three (\textit{i.e.} number of TLVs), the occurrence frequency of OTUs, and lastly the assigned OTU number (ID). Although this last tiebreak is rarely reached, this criterion is necessary to provide a consistent and unique solution to the problem as it will always provide a consistent tiebreak solution due to the uniqueness of the ID. Lower IDs take precedence over higher IDs because it is assumed that in a growing database with data of several contributing international studies, the more common OTUs are sampled first and will have lower ID than the subsequent studies that will add more OTUs to the database. These rules define a partial order on the set of edges between vertices~\cite{Francisco2009} and therefore goeBURST can be considered a distance-based MST algorithm.

goeBURST scheme is very similar to Algorithm~\ref{alg:hc-scheme} but skips the reduction step. In order to build a final optimal forest the algorithm starts with a forest of singleton trees (where each OTU is a tree) and then, iteratively, selects links connecting those in different trees with the minimum distance. This corresponds to selecting a pair of OTUs with maximum number of SLVs at step 1 and joining them at step 2. The final step is not necessary because it does not need to update the overall pairwise distances. goeBURST will always provide an optimal solution for the link assignment, since it performs a global optimization taking into consideration all possible ties at all levels between STs in the data set. 

The final phylogenetic tree is graph-based because topologies like dendrograms frequently do not depict the precise evolutionary history and are particularly susceptible to the confounding effects of recombination. This kind of tree representation may more easily lead to false conclusions about the relationships between species whereas the graph-based tree representation should be preferred when recombination is high~\cite{Achtman2008}.

An example of goeBURST phylogenetic tree with allelic distances of one, two and three are presented in Figure \ref{fig:goeburst} (a), (b) and (c) respectively, and according to distance matrix \ref{matrix:dist}. As one can observe, goeBURST algorithm can construct different forests according to a defined allelic distance.

Using an extension of the goeBURST rules defined above up to $n$LV level (where $n$ equals to the number of loci in a strain), a Minimum Spanning Tree-like structure can be computed. If one define $n$ as one, two or three, the results of this algorithm will be equivalent to calculating goeBURST at those levels (SLV, DLV and TLV respectively).

For the full MST computation the Bor\r {u}vka algorithm is used to build the tree, which is also a greedy algorithm for finding a minimum spanning tree in a graph but with the constraint of having all edge weights distinct~\cite{boruvka1926, Cormen2009}. This algorithm is based on merging disjoint components. 

At the beginning, each vertex is considered as a separate component and in each step it merges every component with some other using strictly the cheapest edge of the given component (every edge must have a unique weight). This way it is guaranteed that no cycle may occur in the spanning tree. At each step this algorithm connects at least half of the currently unconnected components, therefore it is clear that the algorithm terminates in $O(\log V)$ steps. Because each step takes up to $O(E)$ operations to find the cheapest edge for all the components, the asymptotic complexity of Bor\r {u}vka's algorithm is $O(E \log V)$, where $E$ is the number of edges, and $V$ is the number of vertices in the graph $G$.

The constraint of having distinct weights for all edges is not a problem for this algorithm. As mentioned before, in the case of two vertices having the same weight we can apply the tiebreak rules that will always provide a consistent and unique solution.

\begin{figure}
	\centering
	\begin{tikzpicture}
	\node (c)  at (-1,2) {C};
	\node (d) at (-0.5,4) {D};
	\node (e) at (1.5,3.25) {E};
	\node (a) at (4.5,3.25) {A};
	\node (b) at (6,3.25) {B};
	
	\node (1)  at (0.75,3.75) {3};
	\node (1)  at (0.55,2.5) {5};
	\node (1)  at (3,3.5) {6};
	\node (1)  at (5.25,3.5) {2};
	
	\draw (d)-\\(e);
	\draw (c)-\\(e);
	\draw (e)-\\(a);
	\draw (a)-\\(b);
	
	\end{tikzpicture}
	\caption{goeBURST Full MST phylogenetic tree according to matrix \ref{matrix:dist}.} 
	\label{fig:mst-example} 
\end{figure}
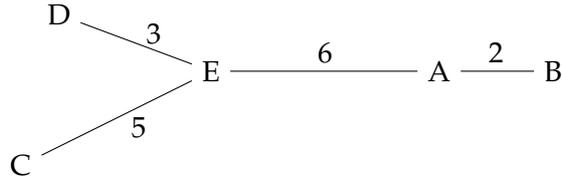

An example of goeBURST Full MST phylogenetic tree according to distance matrix \ref{matrix:dist} is presented in \ref{fig:mst-example}.

\subsection{Properties of reduction formulae} \label{props-reduction-formulae}
All clustering algorithms mentioned before use a specific reduction formula on their last step. In GCP algorithms this is the only thing that distinguish one from another.

Gronau {et al.}~\cite{Gronau2007} demonstrate that if the reduction formula used by these algorithms on step 3 is both convex and commutative, then the selection criterion used on step 1 can be relaxed by joining in each stage a locally (rather than globally) closest cluster-pair. This means that each of the two clusters is closest to the other, but their dissimilarity is not necessarily minimal among all pairwise dissimilarities.
This relaxed selection scheme is called the LCP scheme and was proven to be equivalent (\textit{i.e.} same output tree) to GCP if those two properties of the reduction formula hold.
They also state that this relaxation can be implemented using the Nearest-Neighbor-chain technique where the selection relies on a smaller set (called \textit{NN-chain}) that contains only locally closest neighbors.
We should note that this allows us to speed up several algorithms, such as UPGMA which becomes an $O(n ^2)$ time algorithm,  and it is fundamental in the optimization underlying MST algorithms.

A \textit{convex reduction formula} means that the distance between any cluster $C_k$ to the new cluster $C_u$ (where $C_u=C_i \cup C_j$) lies between the distance from that cluster $C_k$ to  $C_i$ and $C_j$.

A reduction formula is said to be \textit{commutative} if given four arbitrary clusters $\{C_1,C_2,C_3,C_4\}$, the dissimilarity matrix obtained by first joining $\{C_1,C_2\}$ and then joining $\{C_3,C_4\}$ is equal to the dissimilarity matrix obtained by first joining $\{C_3,C_4\}$ and then joining $\{C_1,C_2\}$.

Gronau {et al.}~\cite{Gronau2007} also state that every clustering algorithm that uses a reduction formula which is both convex and commutative can be locally optimized and can be implemented by the NN-chain technique~\cite{murtagh1983survey}. 

Table~\ref{tab:red-form} summarizes these properties for the clustering algorithms mentioned before. We can observe that except GME and BME, they all have a commutative reduction formula and only NJ and its variants cannot assure the convexity property. GME and BME cannot guarantee commutativity because the insertion of a new element in the tree is directly related with the length of the tree after that insertion, which involves changing the length of every other already existing edge. So, changing the order of the addition can lead to different trees. Regarding convexity, NJ and its variants cannot guarantee it because their reduction formula depends on the new branch lengths calculated for the two joined elements and the formula used to compute it do not guarantee convexity, leading to possible negative branch lengths. 

\begin{table*}
\centering
\caption{Local optimization regarding reduction formula properties for different algorithms.}
    \label{tab:red-form}
\begin{tabular}{|c|cc|c|}
\hline
	\multirow{2}{3.5cm}{\centering Algorithm} & \multicolumn{2}{c|}{\multirow{1}{5cm}{\centering Reduction formula}} & \multirow{2}{3.2cm}{\centering Optimized locally}\\
	& \multirow{1}{2cm}{\centering Convex} & \multirow{1}{3cm}{\centering Commutative} & \\\hline
	  	GCP & Yes & Yes & Yes \\
	NJ and variants & No & Yes & No \\
	GME & Yes & No & No \\
 BME & Yes & No & No \\
	goeBURST & Yes & Yes & Yes \\
	Generic distance-based MST & Yes & Yes & Yes \\
	FHP & Yes & Yes & Yes  \\
\hline
\end{tabular}
\end{table*}


\section{Missing data}

The problem of missing data is not uniformly defined in literature,
it varies depending on the definition given on the term "missing data":
it can refer to missing letters, or gaps (whether they are nucleotide or proteins), unsequenced genes or undefined distances between taxa.
We will focus on the problem of missing genes.
Note that missing genes can lead to missing distances in the distance matrix but we can often compute an approximate asymmetric distance.

Intuitively, we could think that adding the genomes with missing data cannot increase the results,
but avoiding them requires us deleting other, probably important, data.
Previously mentioned algorithms should be avoided in their current form as including highly incomplete datasets can often lead to erroneous phylogenetic trees.
Classical MSTs make a lot of mistakes and position incomplete genomes too centrally in a graph.

Some algorithms obtain good results in certain conditions:
Wiens~\cite{wiens2003missing} showed that we can include incomplete datasets in computations without changing the algorithm as long as there is a large amount of complete data.
He ran maximum-likelihood, distanced-based, parsimony and neighbor joining methods on highly incomplete simulated long sequences
(sequences of $2000$ characters with $50\%$ of missing data) and found trees of high accuracy ($>90$\%).
Under conditions closer to reality, algorithms have been created or updated.



Criscuolo and Gascuel~\cite{criscuolo2008fast} updated the criteria of NJ, BIONJ and MVR (Minimum Variance Reduction)~\cite{gascuel2000data} algorithms (updated as NJ*, BIONJ* and MVR*) to allow missing data
(for example: they use the number of existing distances instead of the number of possible distances).
Simulations showed that MVR* is faster than the others and more accurate for high ($75$\%) and low ($25$\%) deletion rates.

An addition to the MST algorithms is the new algorithm of GrapeTree~\cite{zhou2018grapetree}, MSTreeV2, which computes an asymmetric hamming distance, then uses Edmonds' algorithm.


\section{Discussion}
Large epidemiological studies on pathogen populations start to emerge as sequencing technologies become commodity, continuously generating huge volumes of typing data, and also ancillary data.
And there is no doubt about the importance of such studies for the surveillance of infectious diseases and the understanding of pathogen population genetics and evolution.
There are still however a number of challenges.
Phylogenetic analysis is one of the main tools used in this context and, although there are many phylogenetic inference methods as we saw before, their differences and similarities are not clear most of the time.
On the other hand, given the huge volume of ever growing data to analyse, many methods are becoming unpractical due to their computational complexity, even if many of them take polynomial time to run.

We provide in this paper a unifying view on most well known phylogenetic inference methods suitable for processing typing data.
As we discussed, these methods share a common algorithmic background, abstracted here in Algorithm~\ref{alg:hc-scheme}, and differ only on optimization criteria and used heuristics.
Taking this observations into account, one can better understand the difference among these methods and, from a computational point of view, can address simultaneously several challenges in common with all algorithms.
Some of these challenges include the study of how to use approximation schemas or parallelisation to improve methods computational complexity, while controlling the quality of results.

One of the bottlenecks in most studied algorithms is the distance matrix itself.
Since it contains $O(n^2)$ entries, computing it or just parsing it dominates the running time.
This is the case of goeBURST algorithm~\cite{Francisco2009} discussed previously, which runs in subquadratic time
if we are able to avoid computing or parsing the full distance matrix.
Techniques used for approximate string matching can be used to tackle this issue~\cite{carricco2018fast}.
And FastTree2~\cite{price2010fasttree} algorithm also addresses this issue as discussed before in this paper.
Locality-sensitive hashing is yet another technique that allows to avoid the computation
and/or parsing of the distance matrix~\cite{brown2012fast}, allowing to achieve a running time similar to FastTree, but
with a better accuracy.
In the context of phylogenetic tree reconstruction, it makes use of locality-sensitive hash functions
to determine close sequences, estimating where to join tree branches and reconstructing the common ancestor by maximum likelihood.
Overall, this method runs in $O(n^{1+\gamma} \log^2 n)$ time, where $\gamma < 1$ (with $\gamma < 1/2$ for low mutation probability).
All these methods are however sacrificing accuracy for speed and, from an algorithmic perspective,
efficient methods with approximation quality guarantees are of interest and make an interesting research topic.

In spite of these approaches to speedup or avoid the full distance matrix computation, or at least parsing,
given that datasets are not only huge but are growing continuously, dynamic updating is becoming mandatory.
And given the strict relation between clustering and these methods, we believe that well known techniques developed in last years for clustering large data can be useful in this context.

From a practical point of view, distance-based methods presented in this paper
are polynomial time algorithms, making them more usable in practice for large
datasets, with hundreds and even thousands of OTUs, than maximum parsimony or
maximum likelihood methods. We note however that even these polynomial time
algorithms may take long time to run, in particular those with cubic and worst
running times. For even larger datasets, with hundreds of thousands of OTUs, like those
available at EnteroBase (\url{http://enterobase.warwick.ac.uk/}), quadratic algorithms
may be also too slow, and subquadratic ones might be desirable. Although we are trading accuracy
for speed, given that we are dealing with large datasets, that may be a first good approximation.
We can then apply other optimization techniques, such as local optimizations based on local
rearrangements to maximize the likelihood of patterns of descent, as was done for instance in GrapeTree~\cite{zhou2018grapetree}.

Based on the study presented here, we implemented some of these methods within a common framework as open source software and documented at \url{https://gitlab.com/martanascimento/phyloviz/wikis/home}. Furthermore, we list in Table~\ref{tab:ossoft} open source tools that implement most of the discussed methods in this paper.

\begin{sidewaystable}
\centering
\caption{Open source software implementing distance-based methods.}
\label{tab:ossoft}
\begin{tabular}{|c|c|c|c|c|}
\hline
 Distance-Based Methods &  Open Source Software  \\ 
\hline
		 UPGMA~\cite{Sokal1958} & Phylip~\cite{felsenstein2013phylip}; Clustal\cite{chenna2003multiple};  MUSCLE~\cite{edgar2004muscle}; Bosque~\cite{ramirez2008bosque};  PAUP$^{*}$~\cite{swofford2001paup};  SciPy library; PHYLOViZ~\cite{Nascimento2016} \\
		 WPGMA~\cite{Sokal1958} &  SciPy library; PHYLOViZ~\cite{Nascimento2016}\\ 
		SL~\cite{slink} &   SciPy library; PHYLOViZ~\cite{Nascimento2016}\\
		 CL\cite{clink} &   SciPy library; PHYLOViZ~\cite{Nascimento2016}\\
	UPGMC~\cite{sneath1973} & SciPy library\\
		 WPGMC~\cite{sneath1973} & SciPy library\\ 
		 {\multirow{2}{4cm}{\centering NJ~\cite{Saitou1987}} } & Phylip~\cite{felsenstein2013phylip}; RPhylip\cite{revell2014rphylip}; Clustal\cite{chenna2003multiple}; MUSCLE~\cite{edgar2004muscle}; HyPhy~\cite{pond2005hyphy}; \\
 	                                    &  Bosque~\cite{ramirez2008bosque}; MetaPIGA2~\cite{helaers2010metapiga}; 
 T-Rex~\cite{boc2012t};	                                    
 	                                    PHYLOViZ~\cite{Nascimento2016} \\
NJ~\cite{Studier}  &  ape~\cite{paradis2004ape};  QuickTree~\cite{Howe2002}; PAUP$^{*}$~\cite{swofford2001paup};  PHYLOViZ~\cite{Nascimento2016}; FastME~\cite{Lefort2015} \\ 
BIONJ~\cite{GascuelBIONJ1997} & PhyML\cite{guindon2010new}; Bosque~\cite{ramirez2008bosque};   PAUP$^{*}$~\cite{swofford2001paup}; ape~\cite{paradis2004ape};   Phylogeny.fr~\cite{dereeper2008phylogeny}; T-Rex~\cite{boc2012t}; FastME~\cite{Lefort2015}	 \\ 
		UNJ~\cite{gascuel1997concerning} &  T-Rex~\cite{boc2012t}; FastME~\cite{Lefort2015}	 \\ 
		RapidNJ\cite{Simonsen2008} and ERapidNJ~\cite{Simonsen2010} & RapidNJ~\cite{Simonsen2008,Simonsen2010}\\  
	 QuickTree~\cite{Howe2002} & QuickTree~\cite{Howe2002}\\ 
	 QuickJoin~\cite{Mailund2004} &  QuickJoin~\cite{Mailund2004}\\ 
	 NINJA~\cite{Wheeler2009} & NINJA~\cite{Wheeler2009}; T-Rex~\cite{boc2012t} \\ 
	 FastJoin~\cite{Wang2012} &  --- \\ 
	ClearCut (RNJ)~\cite{Sheneman2006} & ClearCut~\cite{Sheneman2006}\\ 
	 FastME~\cite{Lefort2015} &    NRPhylogeny.fr~\cite{lemoine2019ngphylogeny}; FastME~\cite{Lefort2015} \\
	 FastTree~\cite{price2009fasttree},  FastTree 2~\cite{price2010fasttree} & FastTree~\cite{price2009fasttree,price2010fasttree}; NRPhylogeny.fr~\cite{lemoine2019ngphylogeny}\\
	 FNJ~\cite{Elias2009} &  Fastphylo\cite{khan2013fastphylo}\\ 
     GME~\cite{Desper2002}  &  FastME~\cite{Lefort2015} \\ 
	BME~\cite{Desper2002}&    FastME~\cite{Lefort2015}  \\
	 goeBURST~\cite{Francisco2009} &  PHYLOViZ~\cite{Nascimento2016}\\
     FHP~\cite{foulds1979}  &   --- \\		
      NJ*~\cite{criscuolo2008fast} & PhyD*~\cite{criscuolo2008fast}; ape~\cite{paradis2004ape} \\
       BIONJ*~\cite{criscuolo2008fast}  & PhyD*~\cite{criscuolo2008fast}; ape~\cite{paradis2004ape} \\
        MVR*~\cite{criscuolo2008fast} & PhyD*~\cite{criscuolo2008fast}; ape~\cite{paradis2004ape} \\
        MSTReeV2~\cite{zhou2018grapetree} & GrapeTree~\cite{zhou2018grapetree} \\
     
\hline
\end{tabular}
\end{sidewaystable}

%
%


\section*{Funding}

This work was partially funded by
Funda\c{c}\~{a}o para a Ci\^{e}ncia e a Tecnologia (FCT) [grants FCT
PTDC/CCI-BIO/29676/2017, TUBITAK/0004/2014, CMUP-ERI/TIC/0046/2014, SAICTPAC/0021/2015, UIDB/50021/2020].


\bibliographystyle{abbrv}
\bibliography{survey}
%

\end{document}